\documentclass[runningheads]{llncs}
\usepackage{graphicx}
\usepackage[export]{adjustbox}
\usepackage{array,xcolor,colortbl}
\usepackage{multirow}
\usepackage[outdir=./]{epstopdf}
\usepackage[export]{adjustbox}
\usepackage{siunitx}
\usepackage{array}
\usepackage{subfig}
\usepackage{color, colortbl}
\usepackage[section]{placeins}
\usepackage{newunicodechar}
\usepackage[utf8]{inputenc}
\usepackage{textcomp}
\usepackage{xcolor}
\usepackage{algorithmic}
\usepackage{pdfpages}
\usepackage{float}
\usepackage{placeins}
\usepackage{comment}

\usepackage{hyperref}

\begin{document}
\title{Leveraged Mel spectrograms using Harmonic and Percussive Components in Speech Emotion Recognition} 

\author{David Hason Rudd\inst{1,*} \and
Huan Huo\inst{1} \and
Guandong Xu\inst{1,2}}
\authorrunning{D. Hason Rudd et al.}
%
\institute{The University of Technology Sydney, 15 Broadway, Ultimo, Australia \and
Data Science Institute, 15 Broadway, Ultimo, Australia\\
\email{\{david.hasonrudd@student, huan.huo, guandong.xu\}@uts.edu.au}\\$*$corresponding author}
\maketitle 
\begin{abstract}
Speech Emotion Recognition (SER) affective technology enables the intelligent embedded devices to interact with sensitivity. Similarly, call centre employees recognise customers' emotions from their pitch, energy, and tone of voice so as to modify their speech for a high-quality interaction with customers. This work explores, for the first time, the effects of the harmonic and percussive components of Mel spectrograms in SER. We attempt to leverage the Mel spectrogram by decomposing distinguishable acoustic features for exploitation in our proposed architecture, which includes a novel feature map generator algorithm, a CNN-based network feature extractor and a multi-layer perceptron (MLP) classifier. This study specifically focuses on effective data augmentation techniques for building an enriched hybrid-based feature map. This process results in a function that outputs a 2D image so that it can be used as input data for a pre-trained CNN-VGG16 feature extractor. Furthermore, we also investigate other acoustic features such as MFCCs, chromagram, spectral contrast, and the tonnetz to assess our proposed framework. A test accuracy of 92.79\% on the Berlin EMO-DB database is achieved. Our result is higher than previous works using CNN-VGG16. 

\end{abstract}
\section{Introduction}
The general motivation of SER systems is to recognize specific features of a speaker's voice in different emotional situations to provide a more personal and often superior user experience~\cite{cowie2001emotion}. For example, a Customer Relationship Management (CRM) team can use SER to determine a customer’s satisfaction by their voice during a call. 
Emotions are universal, although their understandings, interpretations and reflections are particular and partially associated with culture~\cite{alu2017voice}. Unlike speech recognition, there is no standard or integrated approach for recognising emotions and analysing them through human voices~\cite{weninger2015emotion}. 

The fundamental challenge of SER is the extraction of discriminative and robust features from speech signals. Features used for SER are generally categorized as prosodic, acoustic, and linguistic features. The prosodic features include pitch, energy, and zero-crossings of the speech signal~\cite{li2013automatic,meinedo2010age,shriberg2005modeling}. The acoustic features describe speech wave properties including linear predictor coefficients (LPC), mel-scaled power spectrograms (Mel), linear predictor cepstral coefficients (LPCC), power spectral analysis (FFT), power spectrogram chroma (Chroma), and mel-frequency cepstral coefficients (MFCC)~\cite{chu2009environmental}. In SER, the Mel spectrogram, MFCC, and chromagram are the most effective in decoding emotion from a signal~\cite{motlicek2002feature}. 

Among the most common speech feature extraction techniques, this paper addresses a principal question in Emotion Recognition (ER): How can we maximise the advantage of the Mel spectrogram feature to improve SER? This study presents a novel implementation of emotion detection from speech signals by processing harmonic and percussive components of Mel spectrograms and combining the result with the log Mel spectrogram feature. Our primary contribution is the introduction of an effective hybrid acoustic feature map technique that improves SER. First, we employ CNN-VGG16 as a feature extractor 
of emotion identifier, then utilise the MLP networks for classification task. Furthermore, we tune the MLP network parameters using the random search model hyperparameter technique to obtain the best model. 
Based on empirical experiments, we assert that a data augmentation strategy using an efficient prosodic and acoustic feature combination analysis is the key to obtaining state-of-the-art results since input data represents more diversity with enriched features; these characteristics lead to better model generalisation. 

\section{Related works}


Early traditional SER models relied on modification and optimisation of Support Vector Machine (SVM) classifiers to predict emotions such as anger, happiness, and sadness, among others~\cite{rozgic2012ensemble,perez2013utterance,jin2015speech}. Wu et al.~\cite{wu2011automatic} implemented a traditional machine learning method based on EMO-DB~\cite{burkhardt2005database} database. The authors proposed novel sound features named Modulation Spectral Features (MSFs) that combined prosodic features, and they ultimately obtained 85.8\% validation accuracy for speaker-independent classification using a multi-class Linear Discriminant Analysis (LDA) classifier. Similarly, Milton et al.~\cite{milton2013svm} proposed another classical machine learning method for SER by using a combination of three SVMs to classify emotions in the Berlin EMO-DB. Furthermore, Huang et al.~\cite{huang2014speech} introduced a hybrid model called a semi-CNN, which used a deep CNN to learn feature maps and a classic machine learning SVM to classify seven emotions from EMO-DB. The authors utilised spectrograms as the input for their proposed model and achieved 88.3\% and 85.2\% test accuracy for speaker-dependent and speaker-independent classification, respectively. 

The idea of exploiting pre-trained CNN image classifiers~\cite{cummins2017image} for other tasks involves leveraging transfer learning methods in SER. Surprisingly, using speech-based spectrograms as the input images for pre-trained image classifiers produced competitive results when compared with other well-known traditional methods. Badshah et al.~\cite{badshah2017speech} extracted spectrogram speech features, which were then visualised in 2D images and passed to a CNN; this approach achieved a 52\% test accuracy on EMO-DB. 
Demircan and Kahramanli~\cite{demircan2018application} developed several different classifiers 
and obtained test accuracies 92.86\%, 92.86\%, and 90\%, respectively on SVM, KNN and ANN. Additionally, Wang et al.~\cite{wang2015speech} worked on MFCCs feature and proposed an acoustic feature called the Fourier Parameter (FP) 
, which obtained 73.3\% average accuracy with an SVM classifier.
Furthermore, many similar studies were conducted on different databases. Popova et al.~\cite{popova2017emotion} used a fine-tuned DNN and CNN-VGG16 classifier to extract the Mel spectrogram features in the RAVDESS dataset
~\cite{livingstone2018ryerson} and obtained an accuracy of 71\% ~\cite{popova2017emotion}.
Satt et al.~\cite{satt2017efficient} presented another multi-modal LSTM-CNN and proposed a novel feature extraction method based on the paralingual data from spectrograms. The authors obtained 68\% accuracy on the IMOCAP~\cite{busso2008iemocap} database.

In recent years, some works proposed the use of hybrid feature map techniques as input data for CNN-based networks. Meng et al.~\cite{meng2019speech} proposed a feature extraction strategy for Log-Mel spectrograms that extracted a 3D voice feature representation map by combining log Mel spectrograms with the first and second derivatives of the log MelSpec of the raw speech signal. The authors proposed a CNN with a multimodal dilated architecture that used a residual block and BiLSTM (ADRNN) to improve the classifier accuracy. In addition, the ADRNN further enhanced the extraction of speech features using the proposed attention mechanism approach. The model achieved a remarkable performance of 74.96\% and the 90.78\% accuracy of the IEMOCAP and EMO-DB databases. On the other hand, Hajarolasvadi et al. introduced a 3D feature frame technique for use as input data to the network by extracting an 88-dimensional vector of voice features including MFCCs, intensity, and pitch. The model can reduce speech signal feature frames by applying k-means clustering on the extracted features and selecting the k most discriminant frames as keyframes. Then, the feature data placed in the keyframe sequence were encapsulated in a 3D tensor, which produced a final extracted feature map for use as input data for a 3D-CNN-based classifier that used the 10-fold cross-validation method. The authors achieved a weighted accuracy of 72.21\% on EMO-DB. Zhao et al.~\cite{zhao2019speech} proposed a multi-modal 2D CNN-LSTM network and extracted the log of the Mel-spectrograms from the speech signals for use as input data. The outcome of their work is state-of-the-art with the accuracy of 95.89\% for speaker-independent classification on the Berlin EMO-DB.

\section{Methodology}
This section explains the work procedures used to build the hybrid feature map representation in our model. We compute the average of the signal's harmonic and percussive components and combine the result with the log Mel spectrogram feature. The proposed hybrid feature map method can be generalised with other supervised classifiers to obtain better prediction accuracy. 

\subsection{Proposed hybrid features: Harmonic and Percussive components of Mel spectrogram}

\begin{figure}[!h]
\centerline{\includegraphics[scale=0.095]{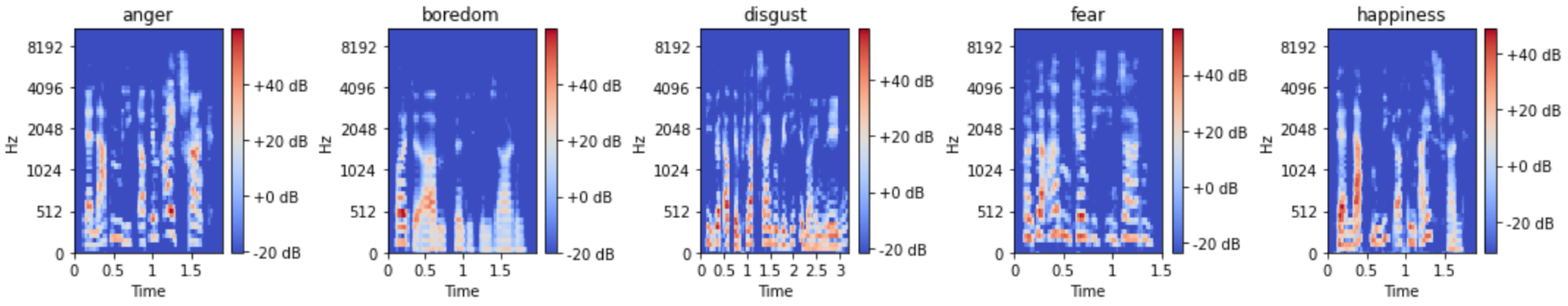}}
\caption{The above sample Mel spectrograms clearly illustrate the distinction between amplitude and frequency in each emotion. The red colours represent frequencies that contribute more than orange and white colours.}
\label{Fig 1}
\end{figure}

Essential features in speech signal processing are the spectrograms on the Mel scale, chromograms~\cite{harte2006detecting}, spectral contrast, the tonnetz~\cite{harte2006detecting} and MFCCs~\cite{xu2004hmm}. 
Since the average length of the recorded voice samples are four seconds, we digitise each original utterance signal at an 88KHz sample rate using the Hanning window function~\cite{harris1978use} shown in (1) to provide sufficient frequency resolution and spectral leakage protection. Next, we apply Mel filter banks to the spectrogram by shifting 0.4ms in a window time of 23ms so that the output is a group of FFTs located next to one another. The Hanning window is described in (1),

\begin{equation}
H_m (n) = 0.5 [1-cos(\frac{2\pi.n}{M-1})] = sin^{2}(\frac{\pi.n}{M-1}) \ \ \\ \ \ \ \ \ 0 \le\ n\ <\ M-1 
\end{equation}

where M represents the number of points in the output window, which is set to 128 and n denote the number of specific sample point from the signal. Finally, we construct the Mel spectrogram by multiplying the obtained energy matrix of the Mel scaled static with the STFT results formulated in (2),
\begin{equation}
LMS(m) = \displaystyle\sum_{k=f(m-1)}^{f(m+1)} log(H_m(k)\ \ .\ \ |X(k)|^{2})
\end{equation}

where $|X(k)|^{2}$ represents the energy spectrum in the $kth$ energy block, $H(k)$ is a Mel-spaced filter bank function, $m$ represents the number of filter banks, and $k$ points to the number of FFT coefficients. $LMS$ represents the log Mel spectrogram. To perform Mel spectrogram feature extraction, we use Librosa tools~\cite{mcfee2015librosa} to set the size of Mel filterbanks as 128, the window size as 2048 and hop length as 512. Fig.~\ref{Fig 1} shows the Mel spectrogram of sample voices exhibiting five emotions from the EMO-DB dataset. It is clear that the amplitude and frequency of each emotion image have a high distinction from other samples.

\begin{figure}[!h]
\centerline{\includegraphics[scale=0.17]{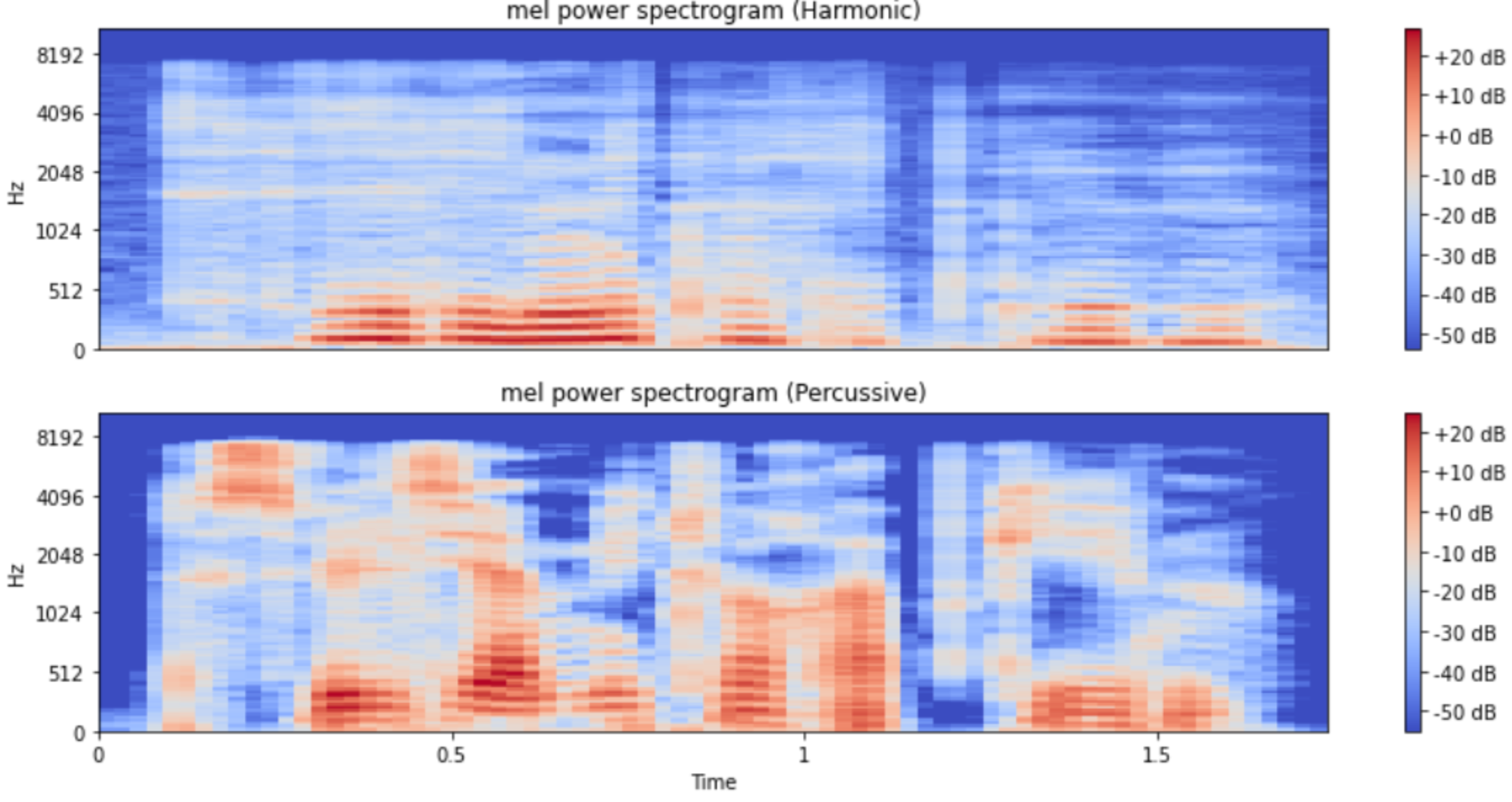}}
\caption{The harmonic and percussive components of the Mel spectrograms for a sample neutral emotion}
\label{fig2}
\end{figure}

The first feature map is built by applying a decomposition process to the Mel spectrum using the popular method in~\cite{fitzgerald2010harmonic}. The decomposition method can be formulated such that the harmonic $s_h$ and percussive $s_p$ components are separated from the input signal $s$ by applying a STFT on the frames to obtain spectrogram $S$ of signal $s$ as shown in definitions (3) and (4),

\begin{equation}
s \ \ = \ \ S_h\ \ + \ \ S_p
\end{equation}

\begin{equation}
S(n,\ k) \ \ := \ \ \displaystyle\sum_{r=0}^{N-1} s(r + nH)\ \ . \ \ \omega(r)\ \ .\ \ e^{(\frac{-j2\pi.k.n}{N}})
\end{equation}

where $S$ denotes a spectrum of signal $s$ in $k^{th}$ Fourier coefficient on the $m^{th}$ time frame, $\omega: [0: N-1]$ := $\{0,1,...,N-1\}$ is a sine windowing function that represents the window length $N$, $H$ represents the hop size value, $n$ indicates current frame number and $N$ is the length of the discrete Fourier transform. We can obtain the harmonic and percussive components of the spectrum by applying a median filter in the horizontal (time-domain) and vertical (frequency-domain) direction on spectrum $S$. Finally, we extract the first feature map by obtaining the mean of both components as shown in the following summarised formulas in (5), (6) and (7),

\begin{equation}
\widehat{H} = \text{\large $\widehat{S}$} \ \ \bigotimes \ \ \text{\large $M$}_H 
\end{equation}
and
\begin{equation}
\widehat{P} = \text{\large $\widehat{S}$} \ \ \bigotimes \ \ \text{\large $M$}_P 
\end{equation}
obtained by
\begin{equation}
\large \mathcal{F}_{2(\text{LMS})} = \frac{\widehat{H} + \widehat{P}}{2}
\end{equation}
where $\bigotimes$ denotes the multiplication element of the median filter in $\text{\large $M$}_H$, which is the horizontal direction filtering used to obtain the $\widehat{H}$ harmonic components of the original spectrogram $\text{\large $\widehat{S}$}$. Subsequently, $\text{\large $M$}_P $ represents the vertical median filtering results $\text{\large $M$}_P$, which is the percussive component of the original spectrogram, $\text{\large $\widehat{S}$}$ shown in (4). Fig.~\ref{fig2} shows the harmonic and percussive components as two distinctive spectrograms in the 128 Mel filterbank.

\begin{figure}[!h]
\centerline{\includegraphics[scale=0.17]{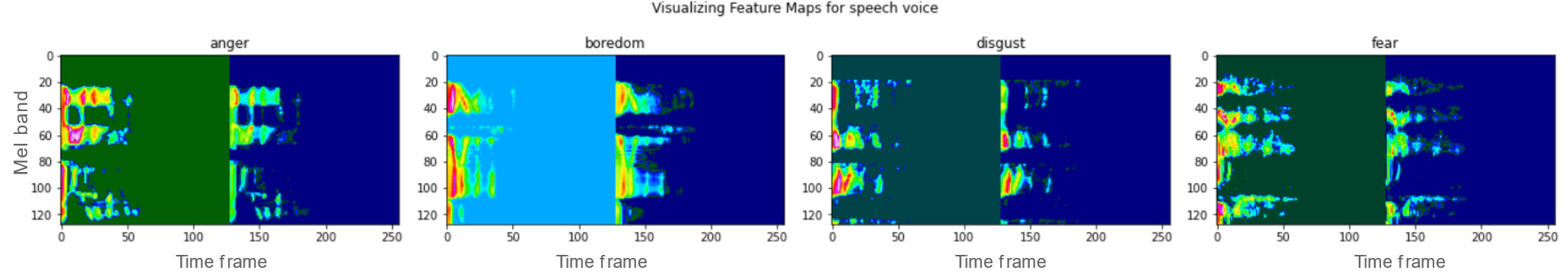}}
\caption{Visualising an achieved 2D hybrid feature maps from the Berlin EMO-DB}
\label{fig3}
\end{figure}

The second feature map is extracted by applying the log of the Mel spectrogram obtained in (2) to measure the sensitivity of the Mel spectrogram output value fluctuation concerning changes in the voice signal amplitude. A sample 2D hybrid feature representation in our work is visualised in Fig.~\ref{fig3}, which clearly shows that each sample feature map is combined in a two-dimensional image. This specific feature combination improves the prediction accuracy in a simple full contact neural network classifier based on our empirical experiments. 

\subsection{Model architecture and training}

We 
use the CNN-VGG16~\cite{russakovsky2015imagenet} as a feature extractor 
to learn from high dimensional feature maps since the network can learn from small variations that occur in the extracted features maps. However, the high-capacity memory storage requirements for a simple classification task can be considered a partial limitation of VGG16 applications.

\begin{figure}[!h]
\centerline{\includegraphics[scale=0.125]{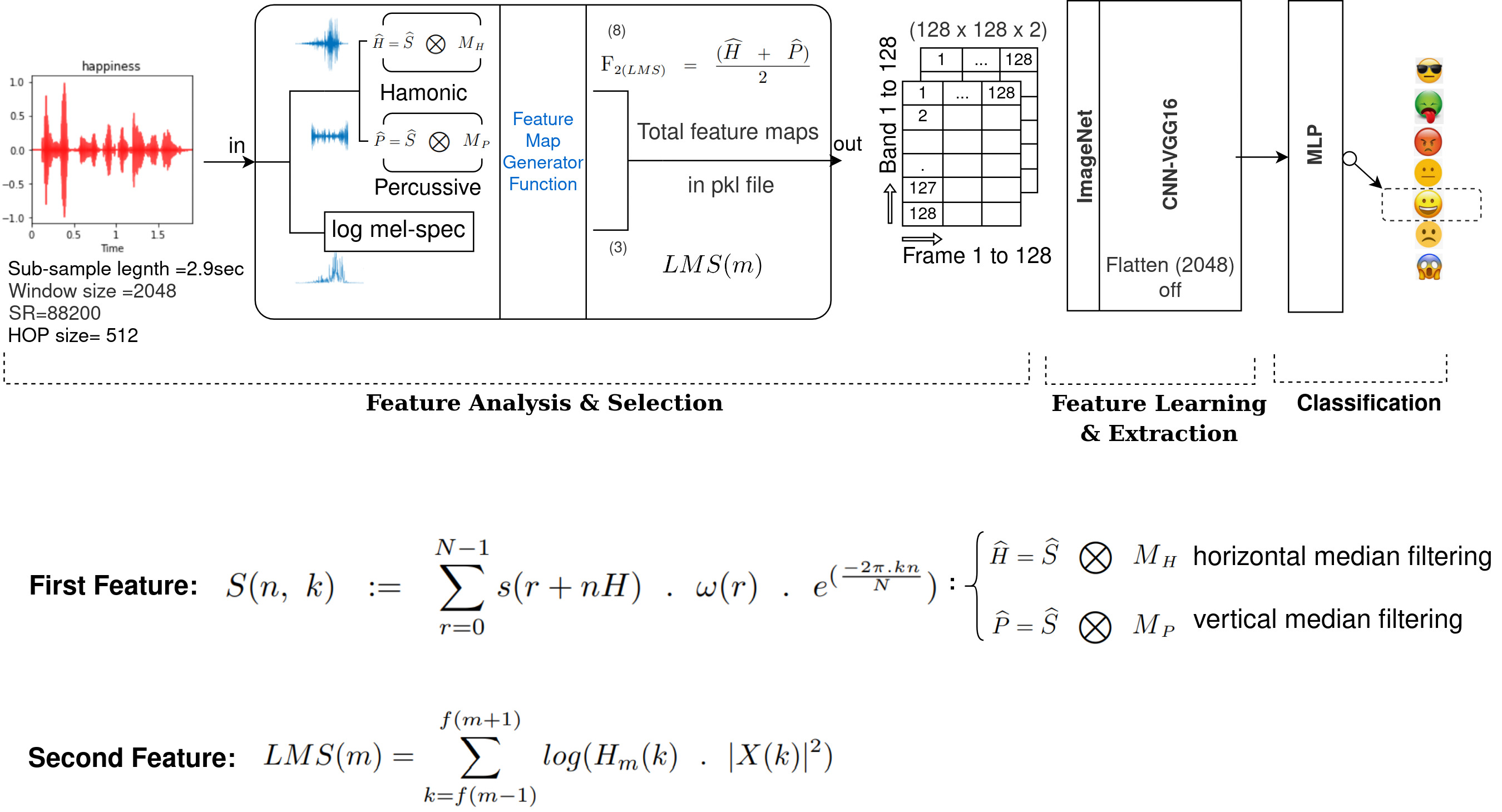}}
\caption{Model architecture, which includes a 2D hybrid feature map built using the harmonic and percussive components, as well as the log of the Mel spectrogram in the feature map generator function. The features are extracted using a CNN-VGG16 network. Finally, the MLP network classifies seven emotions.}
\label{fig4}
\end{figure}

The details of the proposed architecture are shown in Fig.~\ref{fig4}; the architecture consists of an VGG16 and MLP network, which serve as an feature extractor and emotion classifier, respectively.
First, the subsamples are extracted from a fixed window size and then feature maps are built using the proposed feature map function. Therefore, the input to the VGG16 feature extractor is a 2-D feature map in the dimension of (128 x 128 x 2). The input to the MLP classifier is a 2048 one-dimensional vector generated by VGG16. The MLP classifier includes four fully connected layers with the ReLU activation function and softmax in the output layer. Dense 1 and 2 have a 1024 input with a 0.5 dropout value, and dense 3 and 4 are set to 512 input with 0.3 dropouts. The ADAM optimiser with a learning rate of 0.0001 is selected for our architecture design. 

\section {Experimental analysis}
This section analyses the experimental configuration and the result of the feature extractor and MLP classifier 
on EMO-DB~\cite{burkhardt2005database}. The sample voices are randomly partitioned and 80\% are used for the training set and 10\% for the validation and test set for the speaker-independent classification task. We apply an oversampling strategy to compensate the minority classes and increase the voice samples before feeding them to the feature extractor network during the pre-processing phase. The classifier is trained on 128 epochs with a batch size of 128 and used an Nvidia GPU. The window size is set to 2048 with (128 x 128) bands and frames to obtain each subsample length = 2.9 sec. Then, the subsamples are created in each defined data frame. Finally,  167426 signal subsamples and 9717 feature maps are obtained from a sample rate of 88KHz. Based on the time-frequency trade-off, large frame size is chosen to obtain high-frequency resolution rather than time resolution since analysing the frequency of speech signal enables us to decode emotion. 
Several time-consuming experiments are conducted to assess the effectiveness of the proposed hybrid feature, which aims to find the best data augmentation through feature combination. 

\subsection{Results Analysis}
To assess our enriched feature representation method in the MLP classifier, the result of evaluation metrics such as the confusion matrix and test accuracy are observed on different sample rates and feature map dimensions (bands and frames). We also evaluate our model output based on the setting of various parameters in the feature map function. For example, the prediction accuracy results based on some different parameters setting are shown in Table 1. These results indicate that the superior result is achieved on feature map dimensions of 128 x 128 with a sample rate of 88200, Since the highest subsample length of 2.9seconds is achieved and more sample points can contribute in each subsample. 

\begin{table}[h!]
\centering
\caption{The emotion classifier accuracy based on different feature map representation dimensions and signal sampling ratios}
\label{tab:Datasets}
\scriptsize
\begin{tabular}{llllll} \hline
Band \ \ & \ \ Frame \ \ & \ \ Sample rate \ \ & \ \ Accuracy \\ \hline
32 \ \ & \ \ 32 \ \ & \ \ 88200 \ \ & \ \ 71.92\% \\ 
128 \ \ & \ \ 128 \ \ & \ \ 44100 \ \ & \ \ 87.04\% \\ 
128 \ \ & \ \ 128 \ \ & \ \ 22050 \ \ & \ \ 88.54\% \\ 
32 \ \ & \ \ 32 \ \ & \ \ 22050 \ \ & \ \ 89.54\% \\ 
64 \ \ & \ \ 64 \ \ & \ \ 44100 \ \ & \ \ 92.02\% \\ 
128 \ \ & \ \ 128 \ \ & \ \ 88200 \ \ & \ \ \bfseries 92.79\% \\ \hline
\end{tabular} 
\end{table}

We examine the effect of the different number of subsamples from the signal by increasing the window size and sample rate on ten different feature map representations, including 1D, 2D, and 3D maps, and we then compare their results with our hybrid feature extraction method. With respect to the primary research question, it is found that we can take maximum advantage of the powerful Mel spectrogram feature through harmonic and percussive components in emotion recognition. 

As shown in Table 2, the proposed hybrid feature map representation achieves better results than other well-known feature combinations techniques. Furthermore, the results in  Table 2. indicate that the accuracy increases in the high range of the sample rate and window size in most represented methods since the feature map generator function handles more data points via a higher overlapping between frames. Consequently, for most feature extraction methods, the VGG16 network can learn from better-enriched features when the sample rate is higher. In contrast, an increased number of data points in the subsamples requires a memory capacity in the gigabyte range to store the base, train, validation, test feature map files in the pkl format. For instance, in our model, a signal sampling rate of 88KHz and a window size of 2048 occupy an approximately 3-gigabyte memory space to store the pkl files for analysing the whole voice files in the EMO-DB; this requirement can limit its application.  

\begin{table}[h!]
\centering
\caption{ Evaluation of the prediction accuracy based on the different feature extraction methods, sample rate and window size}
\label{tab:Datasets}
\scriptsize
\begin{tabular}{llllll} \hline
Window Size \ \ & \ \ 512 \ \ & \ \ 1024 \ \ & \ \ 2048 \\
Sample Rate \ \ & \ \ 22050 \ \ & \ \ 44100 \ \ & \ \ 88200 \\ 
Feature extraction methods \ \ & \ \ Acc \% \ \ & \ \ Acc \% \ \ & \ \ Acc \% \\ \hline
1D-MFCCs \ \ & \ \ 65.81 \ \ & \ \ 68.39 \ \ & \ \ 69.03 \\
1D-Mel spectrogram \ \ & \ \ 75.48 \ \ & \ \ 75.48 \ \ & \ \ 82.71 \\
1D-Chromagram \ \ & \ \ 80.01 \ \ & \ \ 80.13 \ \ & \ \ 81.29 \\
1D-Tonnets \ \ & \ \ 56.77 \ \ & \ \ 63.08 \ \ & \ \ 56.81 \\
1D-Spectral \ \ & \ \ 54.84 \ \ & \ \ 50.93 \ \ & \ \ 47.10 \\
2D-MFCCs+Chromagram \ \ & \ \ 83.87 \ \ & \ \ 83.23 \ \ & \ \ 91.59 \\
2D-Mel spectrogram+MFCCs \ \ & \ \ 88.39 \ \ & \ \ 85.16 \ \ & \ \ 85.81 \\
2D-Mel spectrogram+Spectral \ \ & \ \ 82.01 \ \ & \ \ 85.13 \ \ & \ \ 80.65 \\
3D-Mel spectrogram+MFCCs+Chromagram \ \ & \ \ 83.87 \ \ & \ \ 88.39 \ \ & \ \ 81.94 \\
2D-log-MSS+Avg.HP(proposed) \ \ & \ \ 92.02 \ \ & \ \ 89.54 \ \ & \ \ \bfseries 92.79 \\ \hline
\end{tabular} 
\end{table}

\begin{figure}[!h]
\centerline{\includegraphics[scale=0.45]{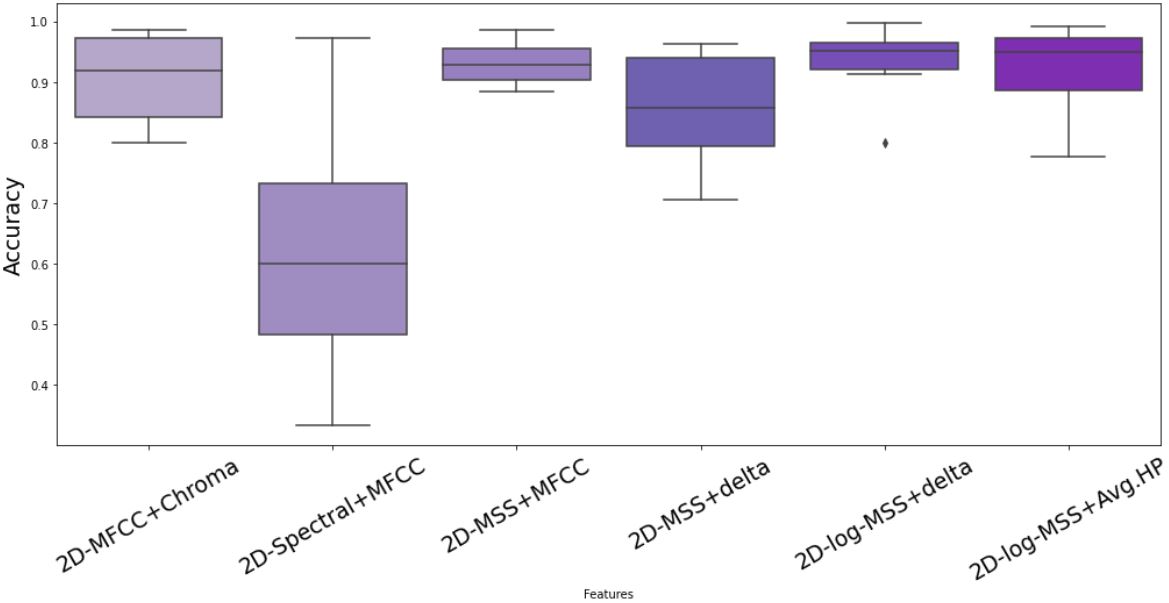}}
\caption{ Variation in the prediction accuracy per emotion class for different feature representation methods}
\label{fig5}
\end{figure}

The fluctuation in the prediction accuracy per emotion class is illustrated for various feature representation methods in Fig.~\ref{fig5}. The boxplot graph shows that the model output is more reliable and stable when predicting seven emotions using our proposed hybrid feature extraction "2D-log-MSS+Avg.HP" and two more feature representations built by combining the delta of the Mel spectrogram (MSS) and log Mel spectrogram or MFCCs features.

The model's confusion matrix in Table 3. shows that the network performs better when recognising specific emotions (anger, sadness, happiness, and fear) while its performance is comparatively poor when predicting emotions such as neutral and boredom. Many experiments are conducted and the highest test accuracy of \%92.79 is achieved.
The Python Keras based network implementation for the proposed model and more experimental results and visualisations are available in our GitHub repositories\footnote{https://github.com/DavidHason/ser}.

\begin{table}[h!]
\centering
\caption{Confusion matrix (\%) of the model with an average accuracy of 92.71\% on the EMO-DB dataset}
\label{tab:Datasets}
\scriptsize
\begin{tabular}{llccccccc} 
\ \ & \ \ \ \ & \ \ \ \ & \ \ \ \ & \ \ \ \ & \ \  \ \ & \ \ \ \ & \ \ \ \ & \ \ \\ \hline
\ \ & \ \ Emotion:  &  Anger  &  Boredom  &  Disgust  &  Fear  &  Happiness  &  Neutral &  Sadness\\ \hline

 & \ \ Anger  & \ \ \bfseries94.92  & \ \ 0  & \ \ 0  & \ \ 0  & \ \ 5.12  & \ \ 0  & \ \ 0 \ \

\\

 & \ \ Boredom  & \ \ 0  & \ \ \bfseries78.77  & \ \ 0  & \ \ 0  & \ \ 0  & \ \ 9.9  & \ \ 11.54 \ \

\\

 & \ \ Disgust  & \ \ 0  & \ \ 0  & \ \ \bfseries89.47  & \ \ 0  & \ \ 9.8  & \ \ 0  & \ \ 0 \ \

\\

 & \ \ Fear  & \ \ 0  & \ \ 0  & \ \ 0  & \ \ \bfseries96  & \ \ 0  & \ \ 0  & \ \ 3.85 \ \

\\

 & \ \ Happiness  & \ \ 0  & \ \ 0  & \ \ 0  & \ \ 0  & \ \ \bfseries100  & \ \ 0  & \ \ 0 \ \

\\

 & \ \ Neutral  & \ \ 0  & \ \ 12.81  & \ \ 0  & \ \ 0  & \ \ 0  & \ \ \bfseries88.87  & \ \ 0 \ \

\\

 & \ \ Sadness  & \ \ 0  & \ \ 0  & \ \ 0  & \ \ 0  & \ \ 0  & \ \ 0  & \ \ \bfseries100 \ \ \\ \hline
\end{tabular} 
\end{table}

\subsection{Model comparison with previous works on EMO-DB}

As shown in Table 4, our method achieves superior results compared with most previous studies except for two works in terms of accuracy that are not significantly higher than our results. However, their work frame is more sophisticated than our proposed model. Zhao et al.~\cite{zhao2019speech} combined two 1-D and 2-D LSTM CNN networks in the feature learning process. Demirican et al.~\cite{demircan2018application} used a model with three classifiers KNN, SVM and ANN to improve the prediction accuracy. Nevertheless, the major advantage of our architecture comes from its simplicity and generality, which can be employed for other acoustic features, as shown in Table 2. Another advantage of the architecture is the capability of storing the feature maps into cloud storage in pkl format that enables us to share them for simultaneous analysis with other networks.

\begin{table}[h!]
\centering
\caption{Model comparison with previous works on EMO-DB}
\label{tab:Datasets}
\scriptsize
\begin{tabular}{llll} \hline
Previous works & Learner &  Feature extraction method &  Accuracy \\ \hline
Badshah et al.~\cite{badshah2017speech}  & CNN & log Mel spectrogram  &  52\% \\ 

Popova et al.~\cite{popova2017emotion} & VGG16 &  Mel spectrograms  &  71\% \\ 

Hajarol. et al.~\cite{hajarolasvadi20193d} & CNN &  Mel spectrograms+MFCCs  &  72.21\% \\ 

Wang et al.~\cite{wang2015speech} & SVM &  Fourier Parameter+MFCCs  &  73.3\% \\ 

Huang et al.~\cite{huang2014speech} & CNN  &  Spectrogram  &  85.2\% \\ 

Issa et al.~\cite{issa2020speech} & VGG16 &  MFCCs+Chroma.+Mel spec.+Contrast+Tonnetz  &  86.10\% \\ 

Meng et al.~\cite{meng2019speech} & CNN-LSTM &  log Mel spec.+1st \& 2nd delta(log Mel spec.)  &  90.78\% \\ 

Wu et al.~\cite{wu2011automatic} & SVM  &  Modulation Spectral Features (MSFs)  &  91.60\% \\

\bfseries Our model & \bfseries VGG16-MLP &  \bfseries Harmonic-Percussive (HP)+log Mel spec.  &  \bfseries 92.79\% \\ 

Demircan et al.~\cite{demircan2018application} & SVM &  LPC+MFCCs  &  92.86\% \\ 

Zhao et al.~\cite{zhao2019speech} & CNN-LSTM &  log Mel spectrogram  &  95.89\% \\ \hline

\end{tabular} 
\end{table}

\section{Conclusion}

The key research question in this study focuses on leveraging Mel spectrogram components in a hybrid-based feature engineering technique as well as proposing a novel acoustic feature extraction method to improve emotion recognition. 
The proposed feature map generator function extracts the harmonic and percussive components by applying a median filter on the horizontal (time-domain) and vertical (frequency-domain) directions of the spectrum, and is implemented with a four-layer MLP classifier to predict emotions in the human voice. 
The performance of the proposed hybrid feature technique is tested on the Berlin EMO-DB and compared with other 1D, 2D, and 3D feature extraction methods. To the best of our knowledge, this is the first study on speech emotion recognition that combines this specific component of the spectrogram. 
The results show that our work significantly outperforms most previous works due to its achievement of a 92.79\% test accuracy which is also a superior result in VGG16 feature learning methods.
In future investigations, facial expression analysis and linguistic features can be embedded into the framework to improve the 
emotion recognition as an acoustic-only method is not constant across different languages and cultures. 

\section*{Acknowledgement:}
This work is partially supported by Australian Research Council under grant number: DP22010371, LE220100078, DP200101374 and LP170100891

%
%

%

%
%
\bibliographystyle{splncs04}

\bibliography{mybibliography}

%






\end{document}